# Three-dimensional polarization imaging of $(Ba,Sr)TiO_3$:MgO composites

Patrick Irvin, Jeremy Levy[†]

*Department of Physics and Astronomy, University of Pittsburgh, Pittsburgh, PA 15260*

Ruyan Guo, Amar Bhalla

*Materials Research Laboratory, Pennsylvania State University, University Park, PA 16802*

The dielectric tuning and loss of $(Ba,Sr)TiO_3$:MgO bulk composites depend strongly on the connectivity and interaction among the two phases. To investigate this relationship, the polar structure and dynamics of these composites is mapped as a function of space and time using a pair of three-dimensional probes: second-harmonic confocal scanning optical microscopy (SH-CSOM), which maps ferroelectric polarization in three dimensions, and time-resolved scanning optical microscopy (TR-CSOM), which maps polarization dynamics along two spatial dimensions and one time dimension. SH-CSOM measurements reveal a high degree of homogeneity within the $(Ba,Sr)TiO_3$ regions, while TR-CSOM measurements indicate that topologically connected regions respond with a spatially uniform phase.

PACS numbers: 77.84.Dy , 42.65.Ky, 77.22.Ej, 77.22.Gm, 77.80.Dj

Over the past decade, a variety of ferroelectric materials (including (Ba,Sr)TiO$_3$, (Pb,Sr)TiO$_3$, SrTiO$_3$, KTaO$_3$, Cd$_2$Nb$_2$O$_7$, *etc*.) have been investigated for frequency-agile microwave electronics (FAME). It is highly desirable that materials selected for such devices have suitable dielectric constants and high tunability figure of merit defined by $\kappa = (\varepsilon_0 - \varepsilon_E)/(\varepsilon_0 \tan \delta)$, where $\tan \delta$ is the dielectric loss, and $\varepsilon_0$ and $\varepsilon_E$ are the dielectric permittivity at zero and *E* electric fields. Many of the materials being investigated possess spatially non-uniform structure and properties, and hence it is important to understand how local ferroelectric properties (*i.e.,* polarization state, local permittivity) correlate with the figure of merit, so that improvements in device performance can be made.

Of all the materials investigated for FAME applications, (Ba,Sr)TiO$_3$ (BST) has received the most attention[1,2,3,4] due to its intrinsic high dielectric tunability and capacity to optimize the figure of merit at a given temperature by changing the Ba/Sr ratio. Most studies of BST for FAME applications have been carried out in thin films.[1,2] For bulk applications in microwave phase shifters, varactors, *etc*., lower dielectric constants than are achievable with BST are required. For this reason, several bulk composites have been developed using BST and MgO.[3,4] The relatively low dielectric constant and loss of MgO helps to reduce the effective dielectric constant and loss of the composite while maintaining respectable tunability. In these composites, the connectivity of the two phases plays an important role in the reproducibility of the properties.

Optical techniques have long been used to image properties of interest in ferroelectric materials. Electro-optic contrast in ferroelectrics has been exploited since the discovery of BaTiO$_3$,[5] and has more recently been refined with advances in laser and optical technologies.[6] Imaging is typically performed by collecting information as a function of two spatial dimensions.

To better understand the properties of bulk materials and composites, adding a third dimension is especially helpful. With second-harmonic confocal scanning optical microscopy (SH-CSOM) it is possible to probe for a ferroelectric response within the bulk of the material. Alternatively, the temporal response of a material can be regarded as a third dimension. Time-resolved confocal scanning optical microscopy (TR-CSOM) substitutes time for the third dimension, enabling polar dynamics to be recorded within a single microwave cycle. Both methods provide important and complementary information about the nature of dielectric tuning in these composite single crystal materials.

Three-dimensional maps of the polar response of single crystal composites of $Ba_{0.6}Sr_{0.4}TiO_3$:MgO are obtained using both SH-CSOM and TR-CSOM. A $(Ba_{0.6},Sr_{0.4})TiO_3$ composition, with $T_c$ close to room temperature, is selected for these studies. Single crystals of BST:MgO are grown by the laser heated pedestal growth (LHPG) technique. The details of the growth experiments are reported elsewhere.[7] The c-axis oriented composite crystal plates are cut and polished from the LHPG grown crystals. Surfaces of the plates are polished using 0.1 μm size alumina powder and annealed at 500° C in order to reduce the mechanical surface strain caused during the polishing.

Figure 1 (a) shows an SEM image of a typical sample used for these experiments. Energy dispersive spectroscopy (EDS) of individual grains shows the chemical constituents of the grain chemistry. Figure 1 (b) reveals negligible intermixing of the BST and MgO components.[8] Silver electrodes are deposited on the surface of the crystal, with a gap spacing of 30 μm.

The second harmonic response of a material is highly sensitive to the breaking of inversion symmetry,[9] and has been exploited in diverse applications, from probing living cells[10] to semiconductor quantum dots.[11] Second harmonic generation has been applied to ferroelectric materials to image ferroelectric domain walls in KTiOPO$_4$,[12] LiNbO$_3$, and BaTiO$_3$;[13] to distinguish between surface and bulk effects in SrTiO$_3$;[14] and to produce three dimensional images in lithium triborate..[15]

Here SH-CSOM is used to collect second harmonic light as a function of three spatial dimensions for the BST:MgO composite. Light from a mode-locked Ti: sapphire laser (wavelength λ=800 nm) passes through a spatial filter and is focused with a high numerical aperture microscope objective (NA=0.8) to a *V*=(0.5 μm x 0.5 μm x 1 μm) region within the sample volume (Figure 2 (a)). The high peak intensity of the ultrashort light pulse (~130 fs pulse duration) produces a strong second-harmonic response in the BST, locally generating light with wavelength λ$_{SH}$ = 400 nm. The microscope objective collects both the reflected IR light and SH light; an ultrafast mirror and shortwave pass filter allow only the frequency doubled light to be detected by a near-ultraviolet sensitive avalanche photodiode (APD). By scanning the laser spot over a volume of sample a three-dimensional representation of the ferroelectric polarization in BST: MgO composite is produced.

A series of SH-CSOM images is shown in Figure 2(b) and (c). The bright regions correspond to areas in which a large amount of SH light is collected, corresponding to volumes in which BST is present. MgO possesses a center of symmetry and therefore no SH light is produced; dark areas therefore correspond to volumes containing MgO. Imaging the SH response of the material allows us to probe the sample at different depths, and therefore produce a three-dimensional polar image of the sample. Figure 2(d) shows a three-dimensional rendering[16] of the second harmonic response, which gives a clearer view of the distinct BST and MgO regions. The solid objects in the rendering are areas where there exists a second harmonic response, namely, regions of BST.

TR-CSOM maps the dynamic response of ferroelectric materials to applied microwave frequency electric fields.[17] A diagram of the experiment is shown in Figure 3(a). The same confocal setup used for SH-CSOM is used to focus 800 nm light pulses from the Ti: sapphire laser to a diffraction-limited spot on the sample surface. A sampled portion of the beam is used to generate a reference for a phase-locked oscillator (PLO). The PLO produces electric fields between 2-4 GHz that have a fixed phase with respect to the laser pulses. By scanning an electrical delay line, the time-resolved images of the electrooptic response of the sample are obtained.

Figure 3 (b-c) show TR-CSOM images at two representative time delays, 100 ps and 400 ps, respectively. The temporal response for two distinct BST grains in the BST:MgO composite is shown in Figure 3(d). Note that the responses of the two grains are approximately out of phase by $180°$. This out of phase relationship indicates that the static polarization of the two domains is approximately antiparallel: the application of an electric field causes the polarization to increase in one region and decrease in the other. The polar response at each spatial location in a time-sequence of TR-CSOM images can be fit to extract the Fourier coefficients of the ferroelectric response.[18] Images of the magnitude and phase of the Fourier coefficients are shown in Figure 3(e-f), respectively. The length scales over which polarization is uniform in magnitude correspond roughly to those imaged by SH-CSOM. Hence, the BST regions appear to be either few- or single-domain, based on the spatial uniformity of the microwave response.

Both SH-CSOM and TR-CSOM images of BST: MgO composite show evidence of distinct BST and MgO regions and the TR-CSOM images illustrate that the ferroelectric polarization behaves nearly uniformly within individual grains at microwave frequencies. In addition, there is negligible microwave dispersion within the grains, in contrast to observations made for BST thin films[18]. While the polarization direction between grains appears random, determined likely by local shape considerations and differential thermal contraction after growth, the MgO appears to have little impact on the microwave properties of the BST, thus fulfilling its requirements as a low-loss, low dielectric constant spacer material.

This work was supported by the National Science Foundation (DMR-0333192).

**Figure Captions**

**Figure 1.** (a) SEM image showing grain distribution of BST (light) and MgO (dark). (b) EDS analysis of MgO and BST grains, showing negligible intermixing. Curve B offset for clarity.

**Figure 2.** (a) Schematic of SH-CSOM technique. (b-c) SH-CSOM images acquired at depths 1.6 µm and 4.0 µm below the surface. The bright regions correspond to areas with strong SH collected, and correspond to the BST grains, while the dark regions correspond to MgO. (d) Three-dimensional rendering of SH response of BST: MgO taken over an 8.8 µm by 8.8 µm by 4.0 µm volume using confocal sectioning. Solid objects are BST while transparent regions are MgO.

**Figure 3.** (a) Diagram of TR-CSOM experiment. (b-c) TR-CSOM images of BST: MgO at two distinct time delays $t_D$ = 100 ps (b) and $t_D$ = 400 ps (c). (d) TR-CSOM curves, S($t$) obtained at two locations indicated by square markers in (b-c), corresponding to two distinct BST grains. The temporal response indicates an out-of-phase relationship between the two grains. (e-f) Spatial maps of the magnitude and phase of the TR-CSOM response.

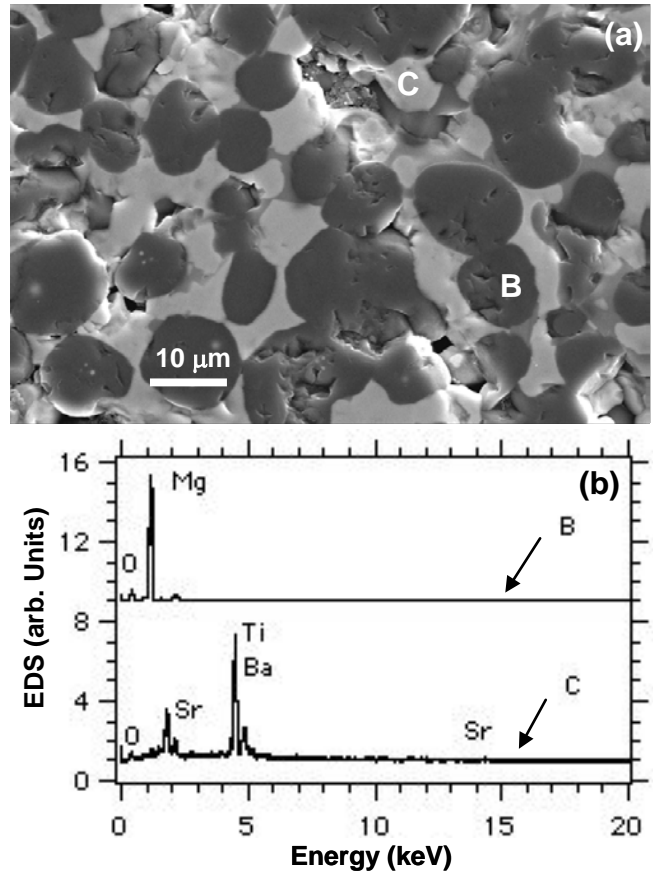

**Figure 2, Patrick Irvin, Applied Physics Letters**

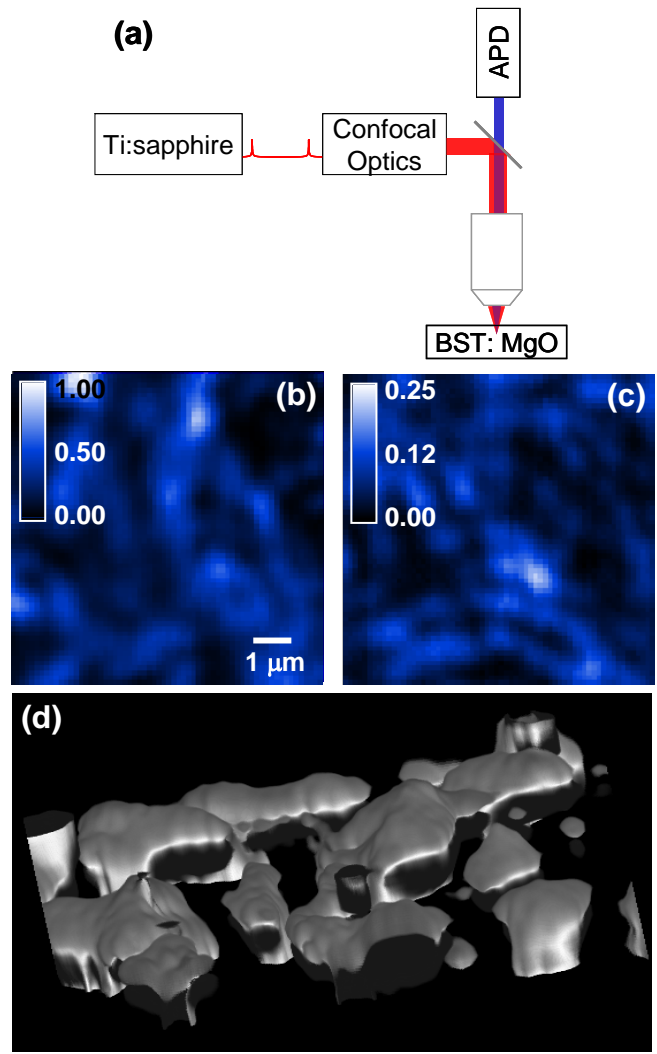

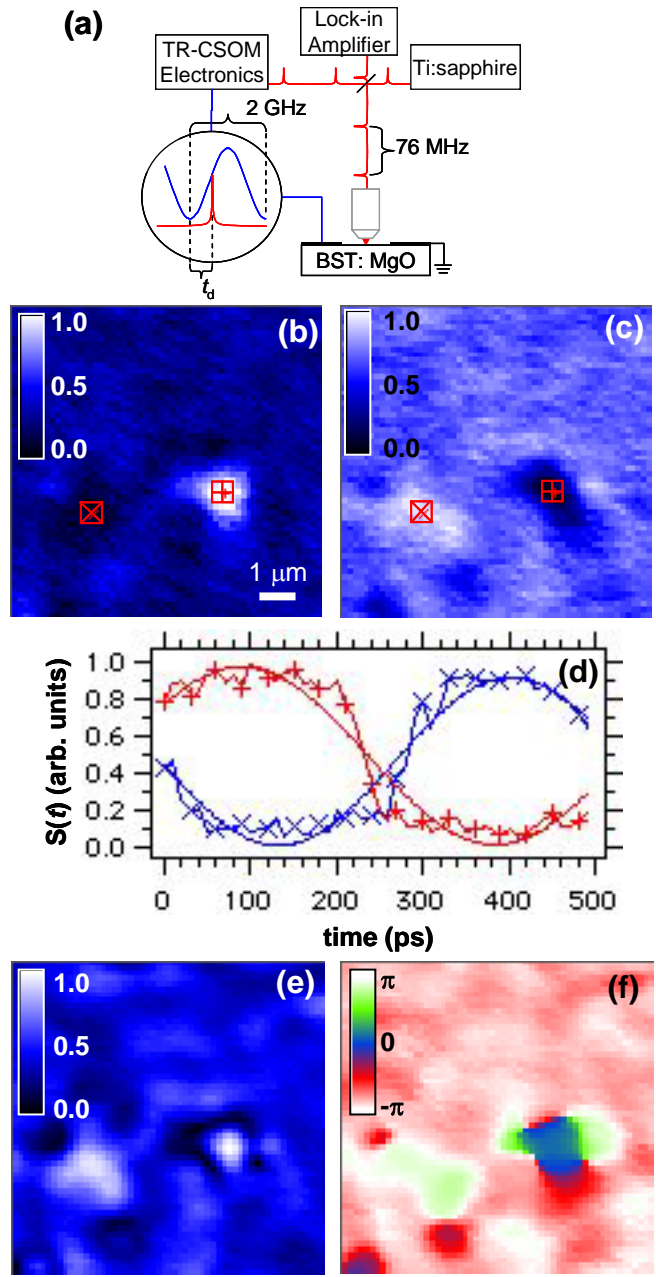